\documentclass[3p,times,procedia]{elsarticle}
\usepackage{nupha_ecrc}
\usepackage{amsmath,bm}
\usepackage{graphicx}
\usepackage{epstopdf}
\usepackage{subfigure}
\usepackage{epsfig}
\usepackage{amsmath,amssymb,amsfonts}
\usepackage{color}
\usepackage[utf8]{inputenc}
\usepackage{verbatim}
\usepackage{array}
\graphicspath{{figures/}} 

\volume{00}

\firstpage{1}

\journalname{Nuclear Physics A}

\runauth{}

\jid{nupha}

\jnltitlelogo{Nuclear Physics A}

\usepackage{amssymb}
\usepackage{amsmath}

\usepackage[figuresright]{rotating}

\begin{document}

\begin{frontmatter}



\dochead{XXVIIIth International Conference on Ultrarelativistic Nucleus-Nucleus Collisions\\ (Quark Matter 2019)}
\dochead{}

\title{Radial distribution of charm quarks in jets in high-energy heavy-ion collisions}

\author[1]{Sa Wang}
\author[2]{Wei Dai}
\author[1]{Jun Yan}
\author[1,3]{Ben-Wei Zhang\footnote{bwzhang@mail.ccnu.edu.cn}}
\author[1,3]{Enke Wang}
\address[1]{Key Laboratory of Quark \& Lepton Physics (MOE) and Institute of Particle Physics,
 Central China Normal University, Wuhan 430079, China}
\address[2]{School of Mathematics and Physics, China University
of Geosciences, Wuhan 430074, China}
\address[3]{Institute of Quantum Matter,
South China Normal University, Guangzhou 510006, China}

\begin{abstract}
Heavy flavor physics in high-energy heavy-ion collisions is a promising and active area to study the mass dependence of the `` jet quenching " effects both at the RHIC and the LHC. In this talk, we present the first theoretical study on the $D^0$ meson radial distributions relative to the jet axis both in p+p and Pb+Pb collisions at $\sqrt{s_{NN}}=5.02$~TeV, where a nice agreement of our results with experimental data is observed. The in-medium parton propagations are described by a Monte Carlo transport model which uses the next-to-leading order (NLO) plus parton shower (PS) event generator SHERPA as input and includes elastic (collisional) and inelastic (radiative) in-medium interaction of heavy flavor jet.  We find that, at low $D^0$ meson $p_T$, the radial distribution significantly shifts to larger radius indicating a strong diffusion effect, and the diffusion effects decrease quickly with $p_T$ ,which is consistent with the recent CMS measurements. We demonstrate that the angular deviation of charm quarks is sensitive to $D_s$ but not $\hat{q}$, which may provide new constrains on the collisional and radiative heavy quark energy loss.

\end{abstract}


\end{frontmatter}


\section{Introduction}
\label{}
In high-energy heavy-ion collisions at the Relativistic Heavy Ion Collider (RHIC) and the Large Hadron Collider (LHC), the deconfied state of quark-gluon plasma (QGP) formed under such extreme hot and dense nuclear matter provides an arena to study the Quantum Chromodynamics (QCD). The strong interaction between the high-$p_T$ jet produced in the hard scattering with the medium, referred as ``jet quenching" effect~\cite{Gyulassy:2003mc}, has been extensively investigated to study the QGP. Especially, owing to the large mass, the early produced heavy quarks are witnesses of the entire bulk medium evolution and strongly interact with the constituents of the QCD matter, therefore being ideal probes to the properties of QGP.

The recent reported $D^0$ meson radial distributions in jets both in p+p and Pb+Pb collisions measured by CMS~\cite{Sirunyan:2019dow} provide new experimental constraints and give new insights into the in-medium interaction mechanisms of heavy quarks inside the quark-gluon plasma. In this talk, we present the first theoretical calculation of $D^0$ radial distribution in jets~\cite{Wang:2019xey} and a systematic discussion on the angluar deviation of charm quarks due to the in-medium interaction.

\section{p+p baseline and in-medium parton energy loss}
\label{baseline}
\begin{figure}[!t]
\begin{center}
\vspace*{-0.1in}
\hspace*{-.1in}
\subfigure[]{
  \epsfig{file=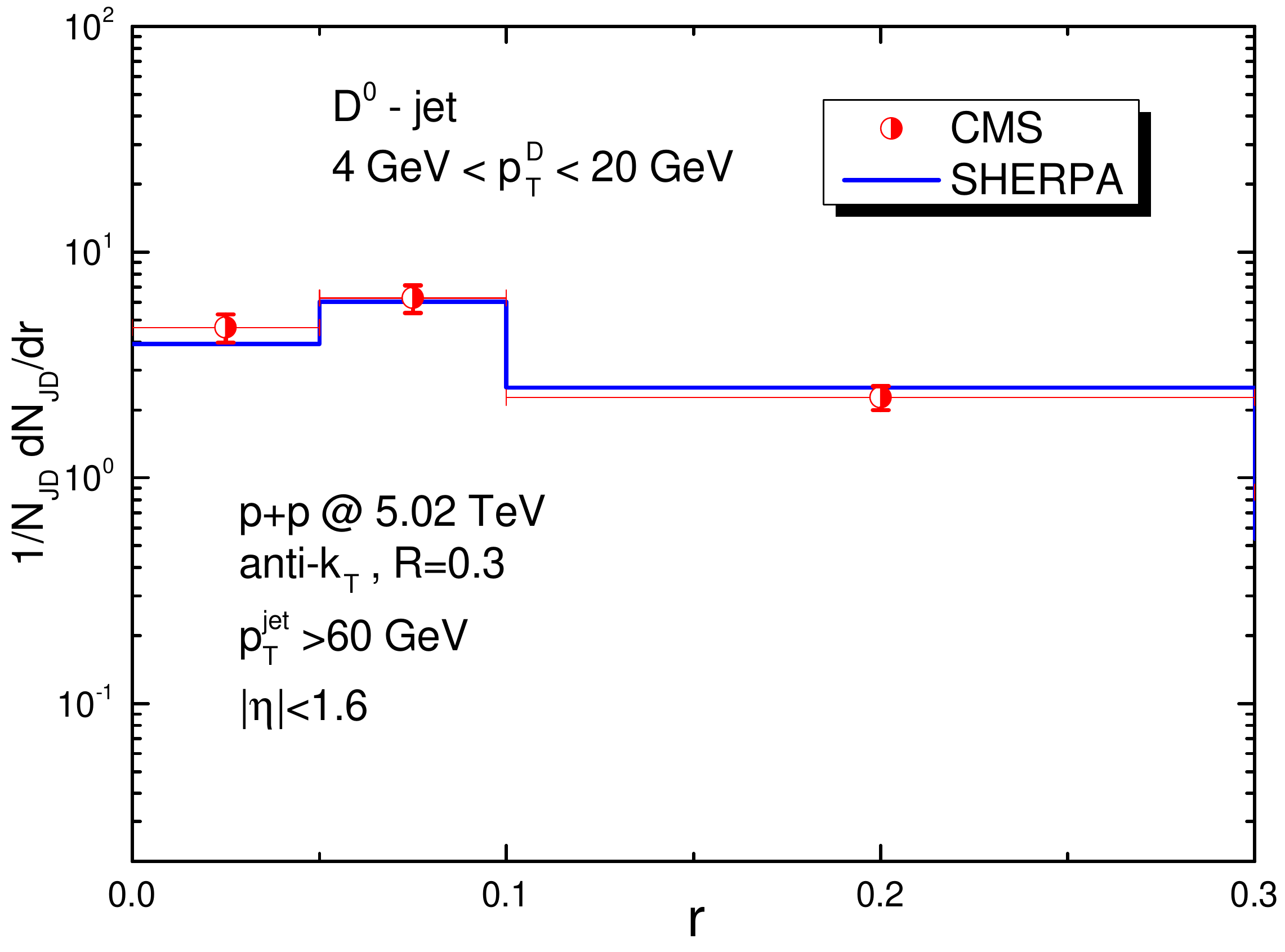, width=2.0in,height=1.8in,angle=0, clip=}}
 \subfigure[]{
  \epsfig{file=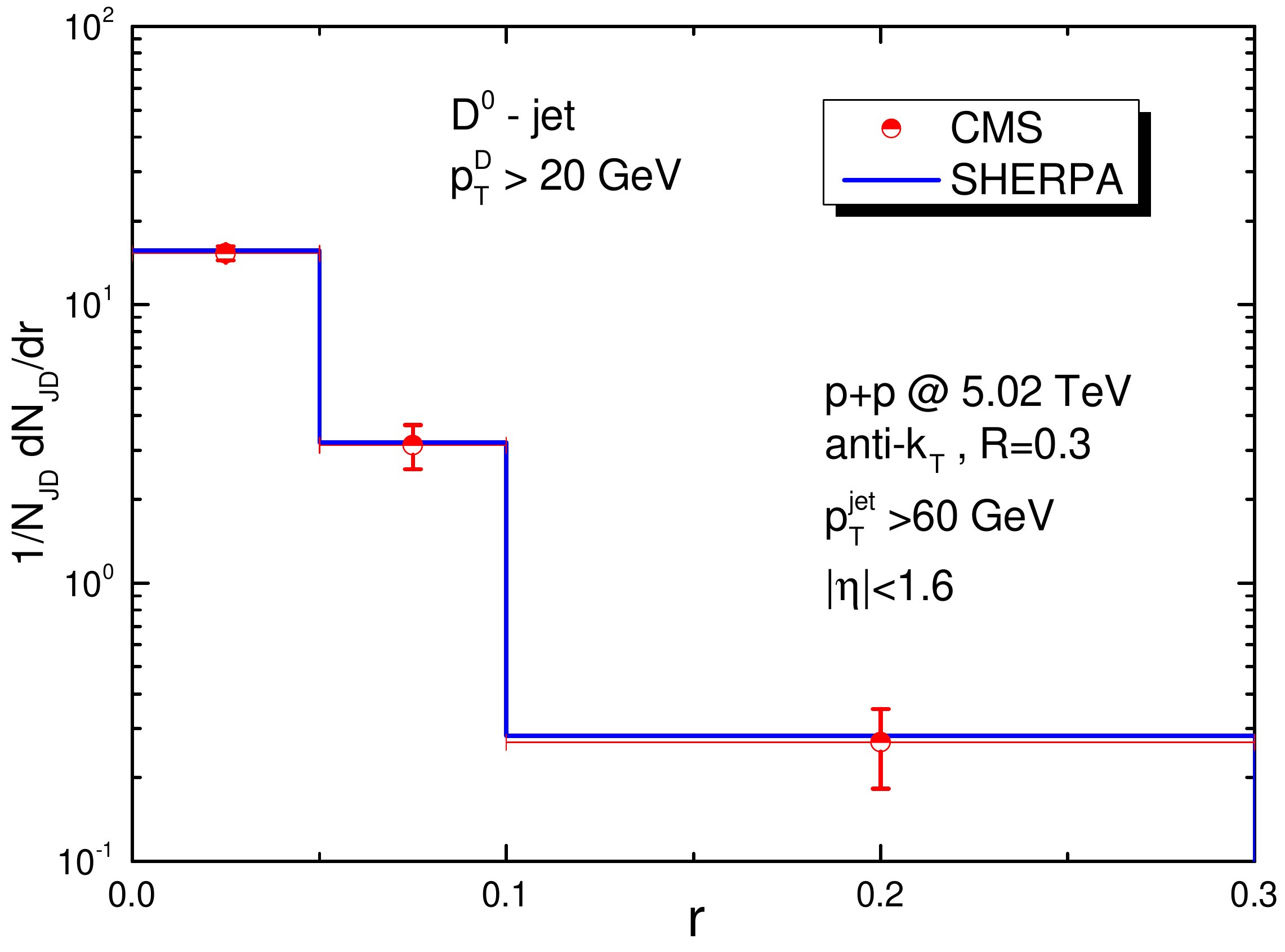,width=2.0in,height=1.8in, clip=}}
\vspace*{-.1in}
\hspace*{0.2in}
\caption{(Color online) The normalized radial distributions of $D^0$ meson in jets as a function of the angular distance from the jet axis in p+p collisions at $5.02$~TeV provided by SHERPA are compared with CMS data at (a) $4$~GeV $< p^D_T < 20$~GeV and (b) $p^D_T > 20$~GeV.}
\label{fig:dndrpp}
\end{center}
\end{figure}

 In this work, we use a Monte Carlo event generator SHERPA~\cite{Gleisberg:2008ta}, which computes the next-to-leading order (NLO) QCD matrix elements matched with parton shower, to generate the initial charm-jet events in p+p collisions at $\sqrt{s_{NN}}=5.02$~TeV. The NNPDF3.0 ~\cite{Ball:2014uwa} parton distribution function (PDF) has been used in our calculation. FastJet~\cite{Cacciari:2011ma} with anti-$k_T$ algorithem was used in jet reconstruction. We use the Peterson form fragmentation functions (FFs) ~\cite{Peterson:1982ak} to perform the hadronization of charm quarks into $D^0$ mesons. Noting that jets are selected at $|\eta^{jet}|<1.6$ and with transverse momentum $p^{jet}_{T}>60$~GeV. In  Fig.~\ref{fig:dndrpp}, we compare the results simulated by SHERPA with the CMS data in two kinematic regions of  $D^0$ meson ( 4~GeV $<p^D_T<$ 20~GeV and $p^D_T>$ 20~GeV), which are in good agreement.
 
To describe the in-medium parton energy loss, elastic and inelastic interactions are usually considered. On one hand, we use a modified Langevin equations incorporating with a recoil term to describe the heavy quark propagation~\cite{Moore:2004tg,Cao:2013ita,Dai:2018mhw,Wang:2018gxz},
\begin{align*}
\vec{x}(t+\Delta t)&=\vec{x}(t)+\frac{\vec{p}(t)}{E}\Delta t \tag{1}\\
\vec{p}(t+\Delta t)&=\vec{p}(t)-\Gamma(p)\vec{p} \Delta t+\vec{\xi}(t)-\vec{p}_g\tag{2}
\end{align*}
where drag coefficient $\Gamma$ and diffusion coefficient $\kappa$ are related by the so called fluctuation-dissipation relation $\kappa=2ET\Gamma=\frac{2T^2}{D_s}$, where $D_s$ denoting the spacial diffusion coefficient is approximatively fixed at $2\pi TD_s=4$ in our simulations based on the Lattice QCD calculation~\cite{Francis:2015daa}. The medium-induced gluon radiation was implemented based on the higher-twist approach~\cite{Guo:2000nz,Zhang:2003yn,Zhang:2003wk,Majumder:2009ge}:
\begin{align*}
\frac{dN}{dxdk^{2}_{\perp}dt}=\frac{2\alpha_{s}C_sP(x)\hat{q}}{\pi k^{4}_{\perp}}\sin^2(\frac{t-t_i}{2\tau_f})(\frac{k^2_{\perp}}{k^2_{\perp}+x^2M^2})^4\tag{3}
\end{align*} 
where $x$ and $k_\perp$ are the energy fraction and transverse momentum carried by the radiated gluon, the last quadruplicate term denote deadcone effect of massive heavy quarks. $\hat{q}$ is the jet transport parameter extracted from a global fitting of single hadron production in A+A collisions~\cite{Ma:2018swx}. The collisional energy loss of light quarks and gluon is performed by the calculation under the Hard Thermal Loop (HTL) approximation~\cite{Neufeld:2010xi}. The hydrodynamic information of the expanding medium is produced by the iEBE-VISHNU model~\cite{Shen:2014vra}.

\section{Results and discussions}
\label{sec:results}

\begin{figure}[!t]
\begin{center}
\vspace*{-0.1in}
\hspace*{-.1in}
\subfigure[]{
  \epsfig{file=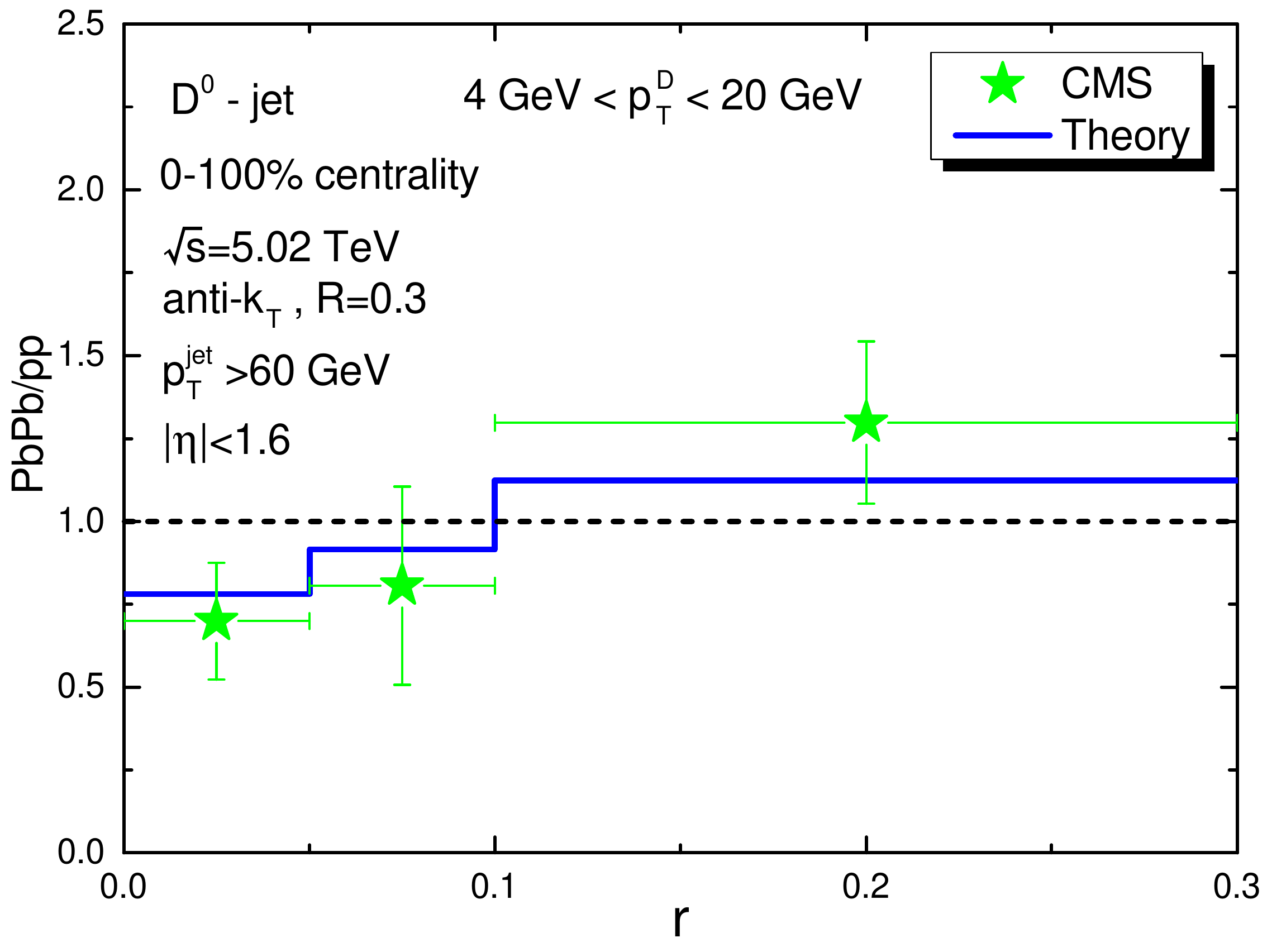, width=2.0in,height=1.8in,angle=0, clip=}}
 \subfigure[]{
  \epsfig{file=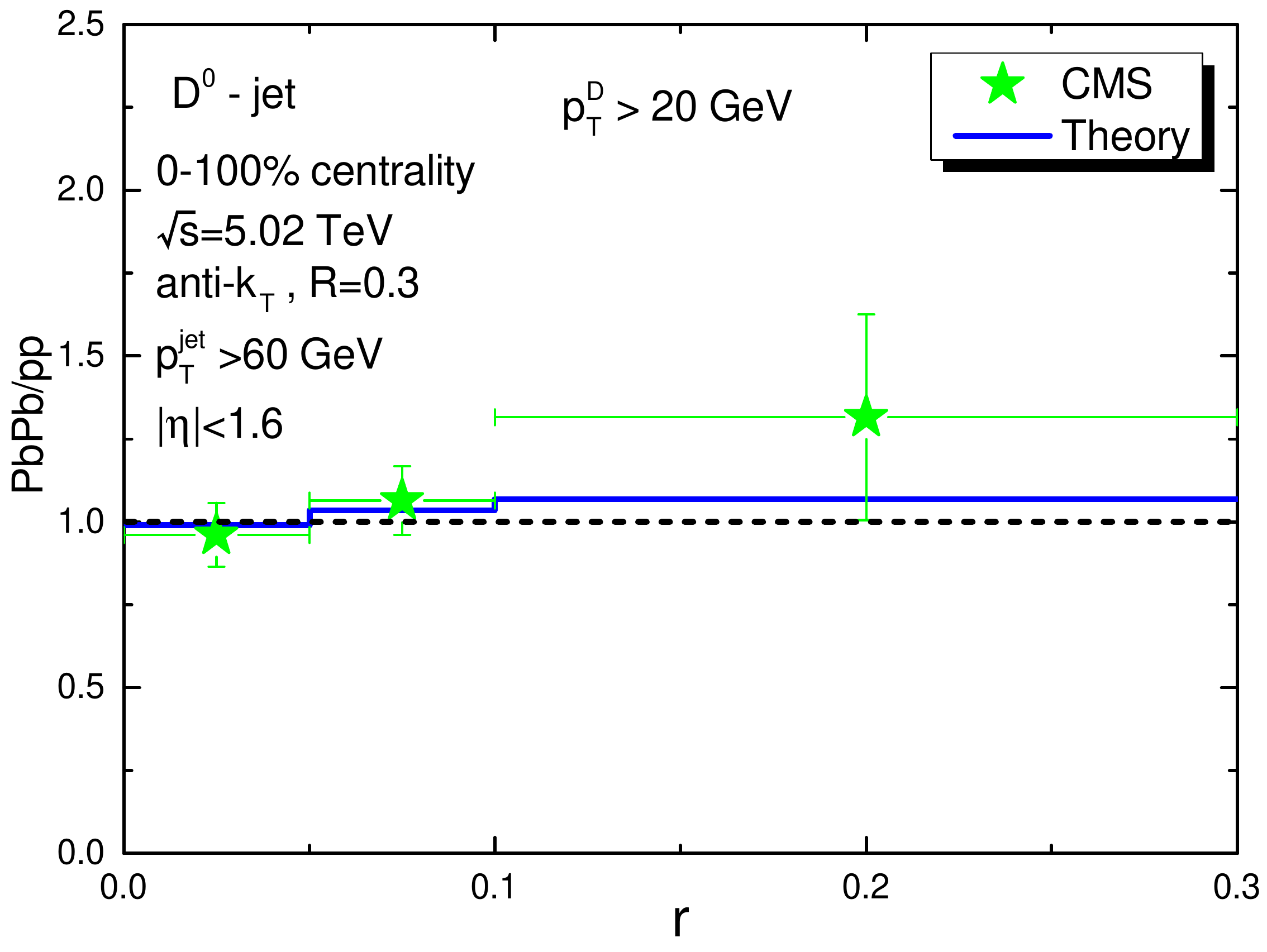,width=2.0in,height=1.8in,angle=0, clip=}}
\vspace*{-.1in}
\hspace*{.2in}
\caption{(Color online) Ratios of the radial distribution of $D^0$ meson in jets of Pb+Pb to p+p in the two $p_T$ range: (a) $4$~GeV $< p^D_T < 20$~GeV and (b) $p^D_T > 20$~GeV are compared with the CMS data~\cite{Sirunyan:2019dow}.}
\label{fig:ratio}
\end{center}
\end{figure}

Within this framework, as shown in Fig.~\ref{fig:ratio}, we calculate the medium modification on the normalized radial distribution of $D^0$ meson in jets at two kinematic region( $4$~GeV$<p^D_T<20$~Gev and $p^D_T>20$~Gev) in Pb+Pb collisions at 5.02 TeV relative to p+p baseline and find that our results give a quiet decent description on the CMS data~\cite{Sirunyan:2019dow}. For low $p_T$ D meson ($4$~GeV$<p^D_T<20$~Gev), suppression near jet axis and enhancement at larger radius indicates that charm quarks diffuse farther away from the jet axis due to the in-medium interaction in Pb+Pb collisions. Whereas, for high $p_T$ $D^0$ mesons ($p^D_T>20$~Gev), the modification is too small. This diffusion effect can be interpreted by two mechanisms: the random kicks by medium constituents in Brown motion and the momentum recoil by medium-induced gluon radiation.

\begin{figure}[!t]
\begin{center}
\vspace*{-0.1in}
\hspace*{.1in}
\subfigure[]{
  \epsfig{file=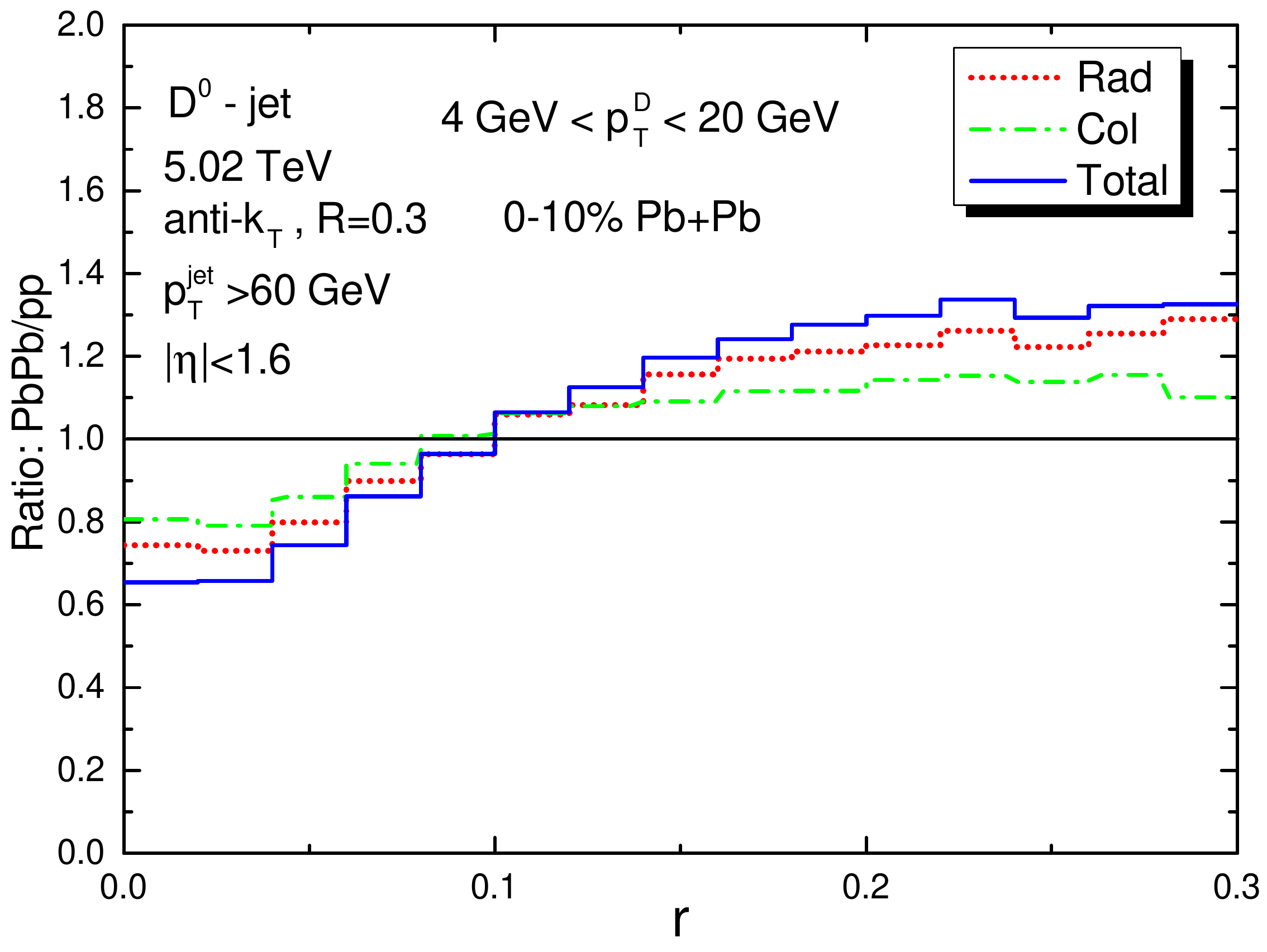, width=1.8in,height=1.6in,angle=0, clip=}}
 \subfigure[]{
  \epsfig{file=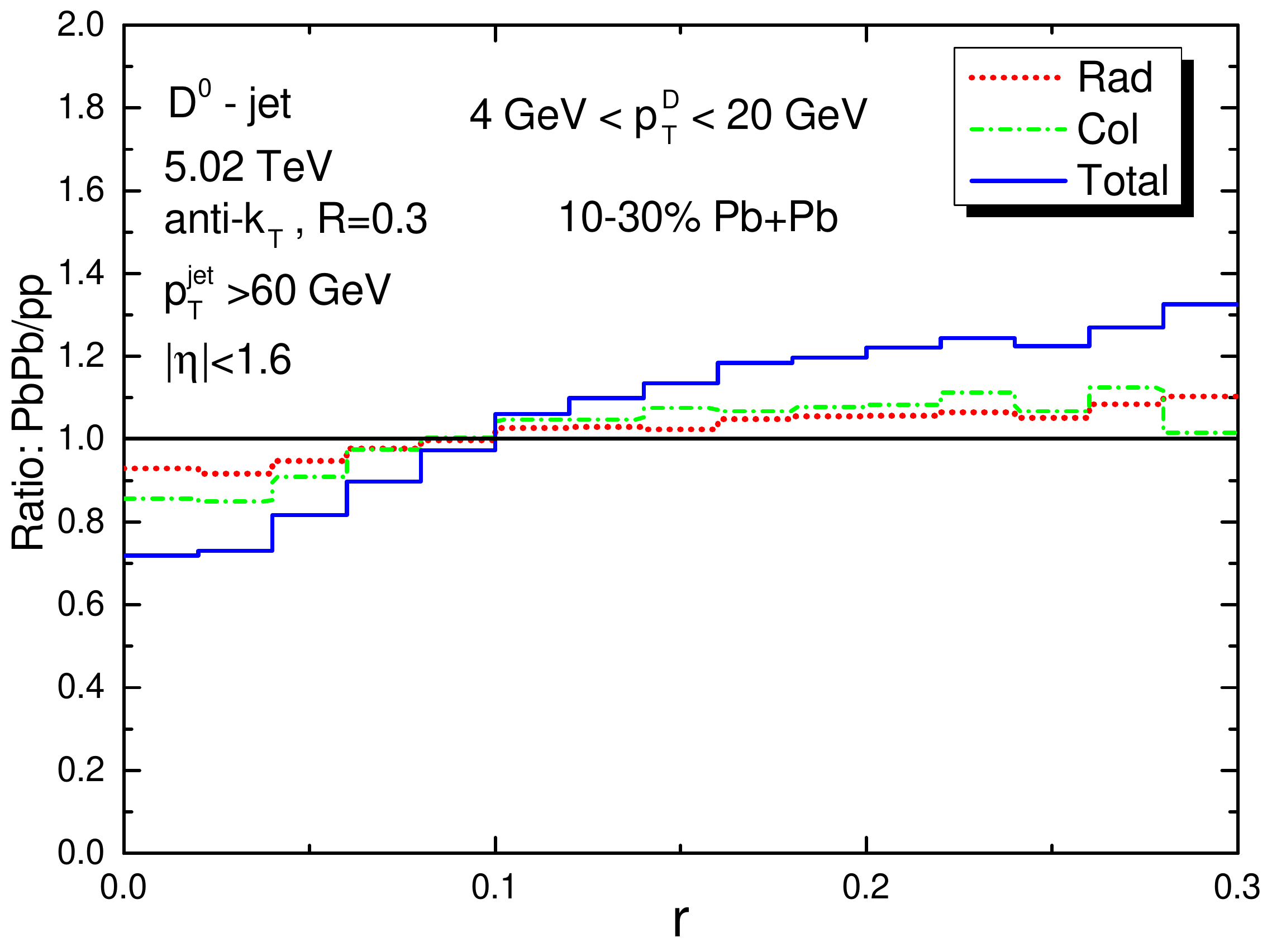,width=1.8in,height=1.6in,angle=0, clip=}}
  \subfigure[]{
  \epsfig{file=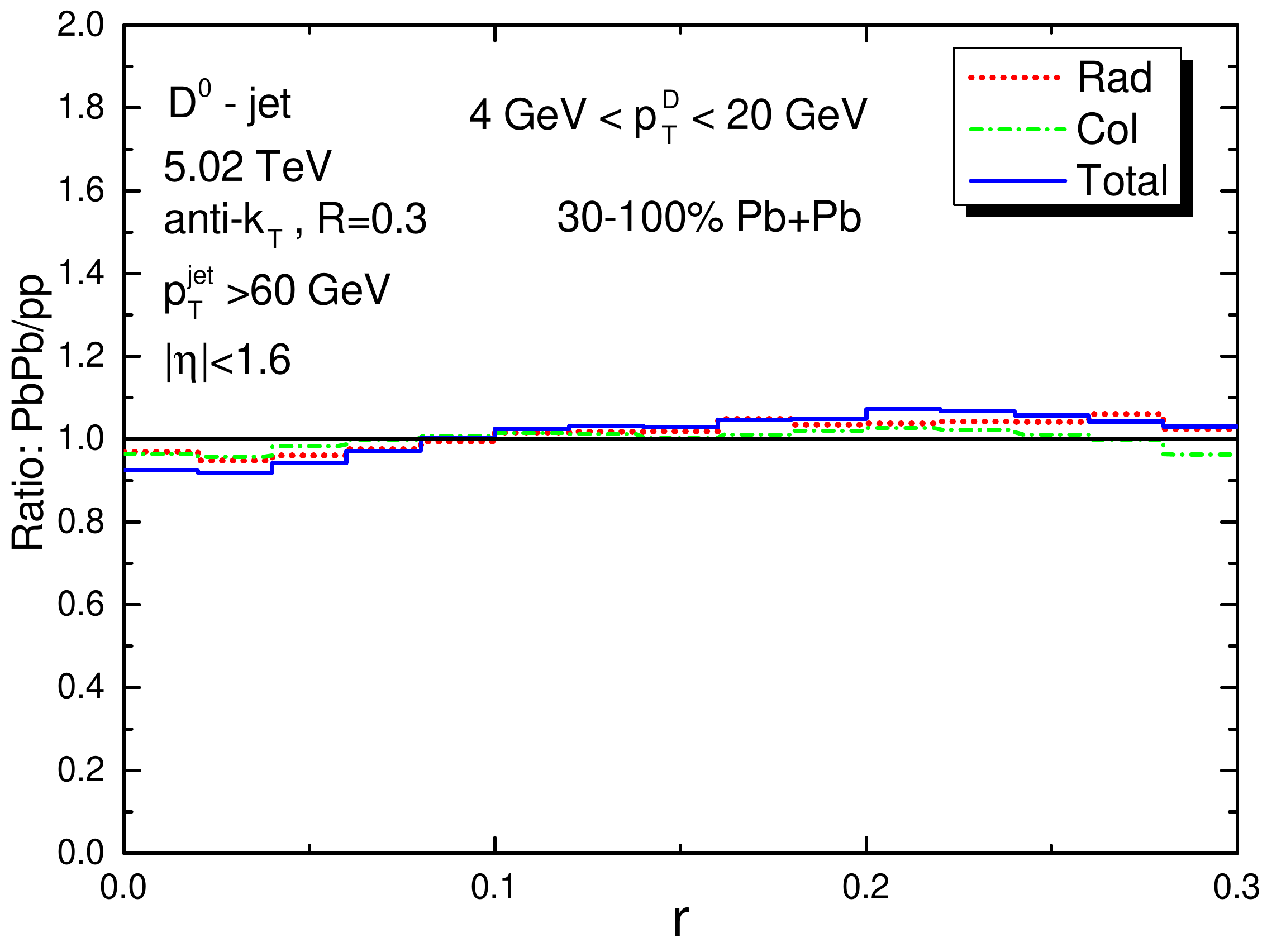,width=1.8in,height=1.6in,angle=0, clip=}}
\vspace*{-0.1in}
\hspace*{.1in}
\caption{(Color online) Ratios of the radial distribution of $D^0$ meson in jets of Pb+Pb to p+p at different centralities : (a) $0-10\%$, (b) $10-30\%$ and (c) $30-100\%$. The net contributions from collisional (green) and radiative (red) interaction are also plot.}
\label{fig:dsdpt-AA}
\end{center}
\end{figure}

Furthermore, shown in Fig.~\ref{fig:dsdpt-AA}, we also estimate the net effect from collisional and radiative interaction on this modification at different centrality: $0-10\%$,$10-30\%$ and $30-100\%$. Obviously, the strongest diffusion effect was obseved in central $0-10\%$ Pb+Pb collisions. This mostly due to the fact that the diffusion effect is sensitive to the medium temperature which depends on collision centrality. In central $0-10\%$ Pb+Pb collisions, we find lager modification due to radiative mechanism than collisional mechanism, but at $10-30\%$ centrality, an opposite situation is observed indiacting that radiative mechanism is more sensitive to collision centraliy than collisional mechanism.

\begin{figure}[!t]
\begin{center}
\vspace*{-0.1in}
\hspace*{.2in}
\subfigure[]{
  \epsfig{file=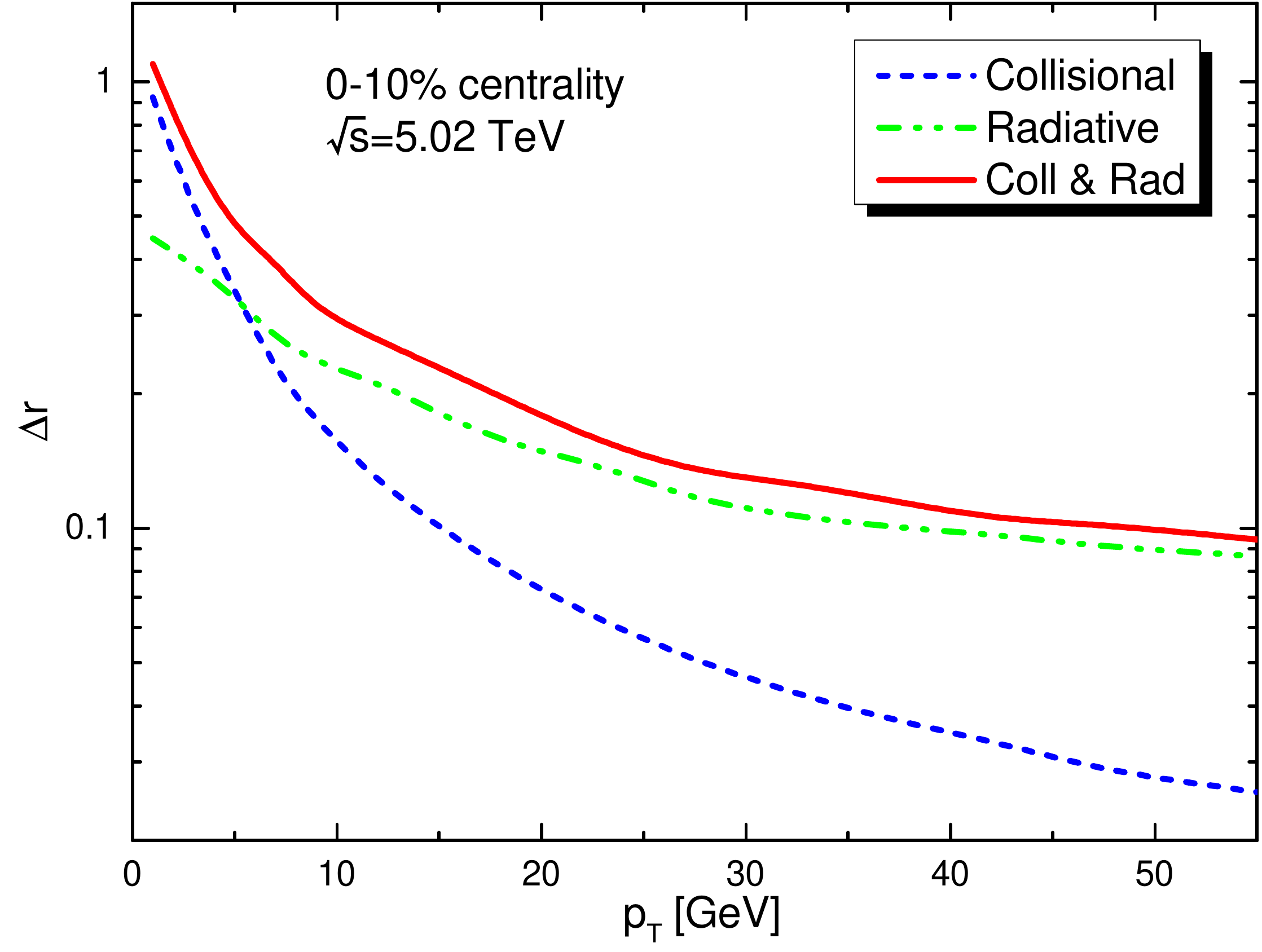, width=2.1in,height=2in,angle=0, clip=}}
 \subfigure[]{
  \epsfig{file=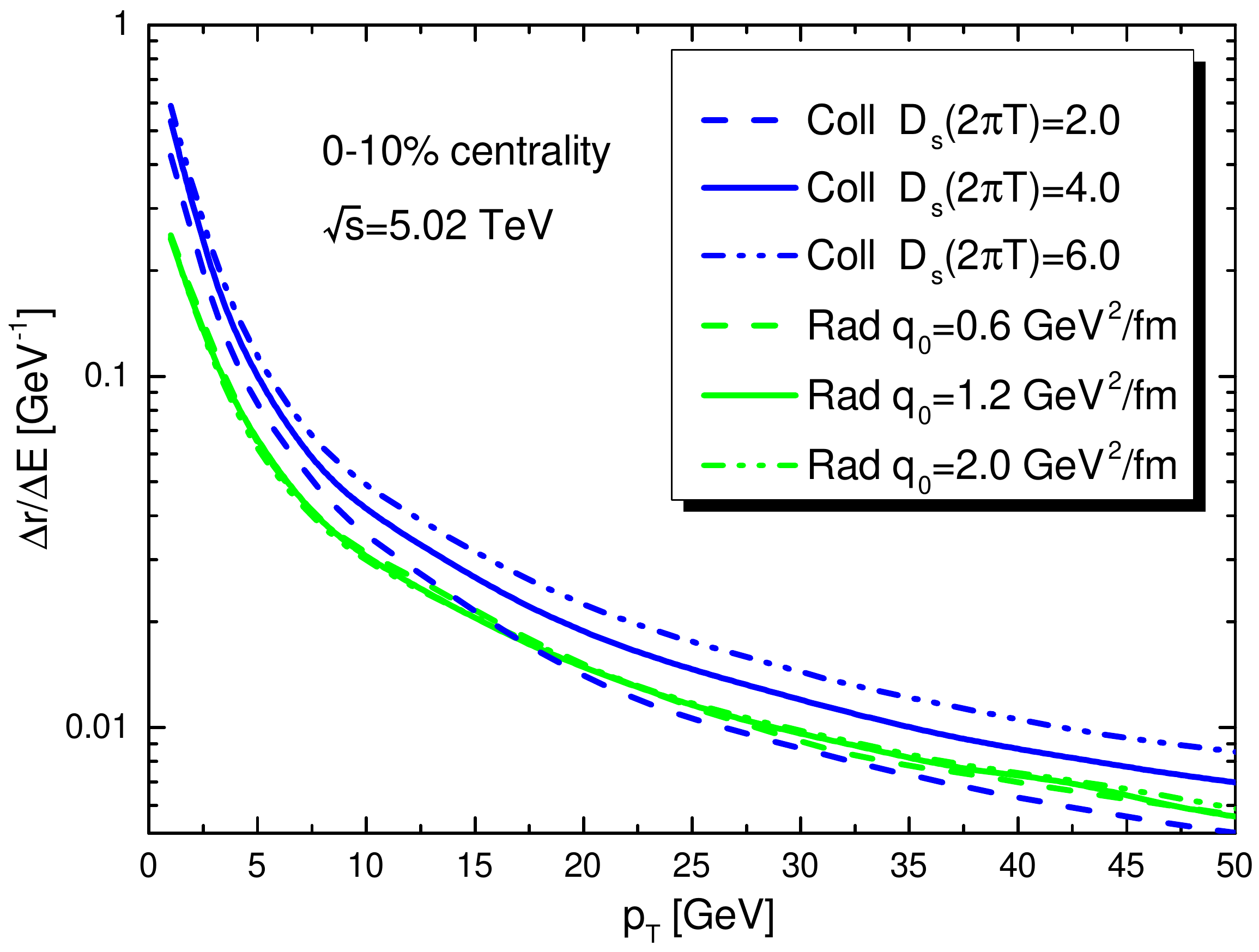,width=2.1in,height=2.05in,angle=0, clip=}}
\vspace*{-.1in}
\hspace*{.2in}
\caption{(Color online) (a) Angular deviation of charm quark in medium as a function of $p_T$ due to elastic interaction (blue) and inelastic interaction (green) as well as the total (red) are shown. (b) Ratio of angular deviation to energy loss versus $p_T$, both for collisional and radiative, by varing the parameter $D_s$ and $\hat{q}$.}
\label{fig:drdt}
\end{center}
\end{figure}

To find out the $p_T$ dependence of this modification observed in experiment and further understand the in-meidum charm diffusion effects result from the elastic and inelastic scarttering, we define $\Delta r\equiv \sqrt{(\phi^c-\phi^c_0)^2+(\eta^c-\eta^c_0)^2}$ to quantify the angular deviation of charm quark from its original moving direction, where $\phi^c_0$ and $\eta^c_0$ are the initial azimuthal angle and pseudorapidity. We show in Fig.~\ref{fig:drdt} (a) the $\Delta r$ as a function of charm quark $p_T$ and find that, at $0-5~GeV$, the diffusion effect is very strong and dominated by collisional process. However, the total diffusion effect falls quickly with increasing $p_T$ which explains the smaller modification of higher $p_T$ $D^0$ meson in Fig.~\ref{fig:ratio} (b), and we find the main contribution is from medium-induced gluon radiation at $p_T>15$~GeV. Beyond that, to estimate the diffusion strength per unit energy loss in these two mechanisms, shown in Fig.~\ref{fig:drdt} (b), we estimate the ratios of angular deviation to the energy loss $\Delta r/\Delta E$ as a function of charm quark $p_T$. We find that $\Delta r/\Delta E$ of collisional interaction is much larger than that of radiative mechanism especially at $p_T < 5$~GeV. In addition to this, we discover that the $\Delta r/\Delta E$ of radiative mechanism is not sensitive to the variance of $q_0$ due to the fact that $\Delta r$ and $\Delta E$ in radiative interaction are all proportional to $\hat{q}$ as implied in the Eq.(3), which is in contrast to the case in collisional interaction : $\Delta r\propto \int \frac{T}{E\sqrt{D_s}}dt$ and $\Delta E \propto \int \frac{T}{D_s}dt$ lead that $\frac{\Delta r}{\Delta E} \propto \sqrt{D_s}$. These different behavior can provide extra constrain on the strength of collisional and radiative interactions. Beyond that, the studies on the radial distribution of b-quark in jets are also under way~\cite{Wang:2020gxz}.

This research is supported by the NSFC of China with Project Nos. 11935007, 11805167, 11435004 and partly supported by China University of Geosciences (Wuhan) (No. 162301182691).




\bibliographystyle{elsarticle-num}
\bibliography{refs}



\end{document}